\documentclass{article}

\usepackage{arxiv}
\usepackage[utf8]{inputenc}
\usepackage[T1]{fontenc}  
\usepackage{hyperref}      
\usepackage{url}         
\usepackage{booktabs}      
\usepackage{amsfonts}      
\usepackage{nicefrac}
\usepackage{microtype}
\usepackage{lipsum}
\usepackage{graphicx}
\usepackage{amsmath}
\usepackage{multirow}
\usepackage{subcaption}
\graphicspath{ {./images/} }

\title{AI-Enhanced Pediatric Pneumonia Detection: A CNN-Based Approach Using Data Augmentation and Generative Adversarial Networks (GANs)}

\author{
 Abdul Manaf \\
  Department of Computer Science\\
  Sukkur IBA University\\
  Sukkur, Pakistan 65200\\
  \texttt{abdul.manaf@iba-suk.edu.pk} \\
   \And
 Nimra Mughal \\
  Department of Computer Science\\
  Sukkur IBA University\\
  Sukkur, Pakistan 65200\\
  \texttt{nimra@iba-suk.edu.pk} \\}

\begin{document}
\maketitle
\begin{abstract}
Pneumonia is a leading cause of mortality in children under five, requiring accurate chest X-ray diagnosis. This study presents a machine learning-based Pediatric Chest Pneumonia Classification System to assist healthcare professionals in diagnosing pneumonia from chest X-ray images. The CNN-based model was trained on 5,863 labeled chest X-ray images from children aged 0-5 years from the Guangzhou Women and Children's Medical Center. To address limited data, we applied augmentation techniques (rotation, zooming, shear, horizontal flipping) and employed GANs to generate synthetic images, addressing class imbalance. The system achieved optimal performance using combined original, augmented, and GAN-generated data, evaluated through accuracy and F1 score metrics. The final model was deployed via a Flask web application, enabling real-time classification with probability estimates. Results demonstrate the potential of deep learning and GANs in improving diagnostic accuracy and efficiency for pediatric pneumonia classification, particularly valuable in resource-limited clinical settings \url{https://github.com/AbdulManaf12/Pediatric-Chest-Pneumonia-Classification}
\end{abstract}


\section{Introduction}
Introduction
Chest X-ray (CXR) imaging represents one of the most widely utilized diagnostic modalities in modern medicine, serving as a cornerstone for evaluating respiratory and cardiovascular conditions \cite{Gambato2023}. This non-invasive imaging technique employs low-dose ionizing radiation to visualize thoracic structures, including the lungs, heart, blood vessels, and skeletal framework \cite{Hussain2022}. The clinical significance of CXR extends across numerous pathological conditions \cite{Speets2006}, encompassing pneumonia, tuberculosis, lung malignancies, and cardiac disorders. In pediatric medicine, CXR serves as an indispensable diagnostic tool for assessing respiratory pathology, enabling healthcare practitioners to rapidly evaluate disease severity and formulate appropriate therapeutic interventions\cite{Sanida2024}.
Despite its widespread clinical utility, CXR interpretation in pediatric populations presents unique diagnostic challenges, particularly in children under five years of age. The interpretation of pediatric chest radiographs is complicated by several factors: the smaller anatomical dimensions of pediatric thoracic structures, less distinct pulmonary parenchymal patterns compared to adults, and the inherent difficulty in maintaining patient immobility during image acquisition. These technical limitations often result in suboptimal image quality, potentially compromising diagnostic accuracy. Furthermore, the clinical presentation of overlapping pathologies, such as concurrent viral and bacterial infections, adds complexity to radiological interpretation, challenging even experienced radiologists in achieving definitive diagnoses.

Artificial Intelligence (AI) encompasses the development of computational systems capable of executing tasks traditionally requiring human cognitive abilities, including decision-making, pattern recognition, problem-solving, and natural language processing. AI systems utilize sophisticated algorithms to analyze vast datasets, extract meaningful patterns, and generate predictions or decisions based on learned representations. Recent years have witnessed exponential growth in AI applications across diverse sectors, including healthcare, finance, education, and autonomous systems \cite{Manaf2025}. Within healthcare, AI technologies are increasingly deployed to enhance diagnostic accuracy, optimize treatment recommendations, and predict patient outcomes.

The healthcare implementation of AI predominantly leverages machine learning (ML), a specialized AI subset focused on improving diagnostic precision and clinical efficiency. ML algorithms are engineered to identify complex patterns within large datasets without explicit programming instructions. In medical imaging applications, deep learning architectures, particularly Convolutional Neural Networks (CNNs), have demonstrated exceptional capability in processing and analyzing radiological images, including X-rays, computed tomography (CT), and magnetic resonance imaging (MRI). These sophisticated algorithms can automatically detect subtle patterns and anomalies that may escape human visual perception. Through training on extensive labeled image datasets, AI systems have achieved remarkable diagnostic accuracy across various pathological conditions, including pneumonia, malignancies, and fractures, often matching or exceeding conventional diagnostic methods.

Multiple research studies have validated the efficacy of AI in chest X-ray interpretation. Developed a deep learning model capable of diagnosing pneumonia from CXR images with performance comparable to practicing radiologists \cite{Rajpurkar2018}. Similarly, Esteva et al. (2017) demonstrated that deep learning algorithms could identify skin cancer lesions with accuracy equivalent to dermatologists\cite{Jeong2023}. These landmark studies underscore AI's transformative potential in enhancing diagnostic precision, reducing clinician workload, and accelerating diagnostic workflows. Continuous improvements in AI-based diagnostic tools, driven by expanding datasets and advancing machine learning methodologies, promise even greater clinical performance.
Pneumonia, particularly in pediatric populations, remains a significant global health burden with substantial mortality implications. Timely and accurate diagnosis is crucial, as bacterial pneumonia requires immediate antibiotic therapy, while viral pneumonia is managed primarily through supportive care. However, pneumonia diagnosis from chest radiographs presents considerable challenges in pediatric patients. The compact anatomical dimensions and developmental variations in pediatric thoracic structures, combined with technical difficulties in obtaining high-quality images, complicate accurate radiological interpretation. Additionally, increasing case volumes often overwhelm radiologists, potentially compromising diagnostic timeliness and accuracy. These challenges necessitate the development of automated classification systems capable of accurately analyzing pediatric chest X-ray images and providing timely diagnostic support for pneumonia detection.

\subsection{Objectives}
The primary objective of this study is to develop a machine learning-based Pediatric Chest Pneumonia Classification System:
\begin{itemize}
  \item Develop an AI-based classification system that can accurately differentiate between normal and pneumonia cases in pediatric chest X-ray images using state-of-the-art deep learning architectures.

  \item Enhance model performance through advanced data augmentation techniques and Generative Adversarial Networks (GANs) to address dataset limitations by generating synthetic images and expanding the training dataset.

  \item Deploy the AI system in a user-friendly web application that provides healthcare professionals with real-time diagnostic support and seamless integration into clinical workflows.
\end{itemize}

\subsection{Research Questions}
This study aims to address the following research questions:
\begin{itemize}
  \item Can deep learning models accurately classify pediatric chest X-ray images to distinguish between normal and pneumonia cases with clinically acceptable precision and recall metrics?

  \item How can data augmentation techniques and GANs be effectively utilized to overcome the inherent challenges of limited medical imaging datasets while maintaining diagnostic accuracy?

  \item What measurable impact does AI-based diagnostic support have on the efficiency and accuracy of pneumonia diagnosis in pediatric chest X-ray interpretation?

  \item Can a web-based deployment of the AI system provide an intuitive and clinically viable interface for healthcare professionals to utilize in real-world clinical settings?
\end{itemize}

\section{Methodology}
This section outlines the methodology employed in developing the Pediatric Chest Pneumonia Classification System, including data collection, preprocessing, model architecture, training, evaluation, and deployment.

\begin{figure}[h]
    \centering
    \includegraphics[width=0.8\textwidth]{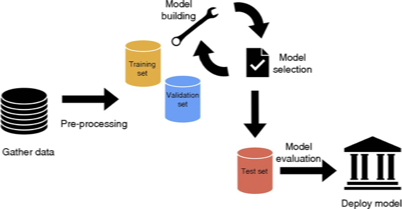}
    \caption{Methodology Flowchart for Pediatric Chest Pneumonia Classification System.}
    \label{fig:methodology_flowchart}
\end{figure}

\subsection{Research Design}
This research adopts a machine learning approach to develop an AI-based system for classifying pediatric chest X-ray images into two categories: normal and pneumonia. The design involves multiple phases, including data collection, preprocessing, augmentation, model training, and evaluation. The system was designed to improve diagnostic accuracy by using deep learning techniques, particularly Convolutional Neural Networks (CNNs) for image classification and Generative Adversarial Networks (GANs) for data augmentation. The research focuses on enhancing model performance by using various techniques, including data augmentation, GAN-generated synthetic images, and fine-tuning hyperparameters.

\subsection{Dataset}
\begin{figure}[h]
    \centering
    \includegraphics[width=0.6\textwidth]{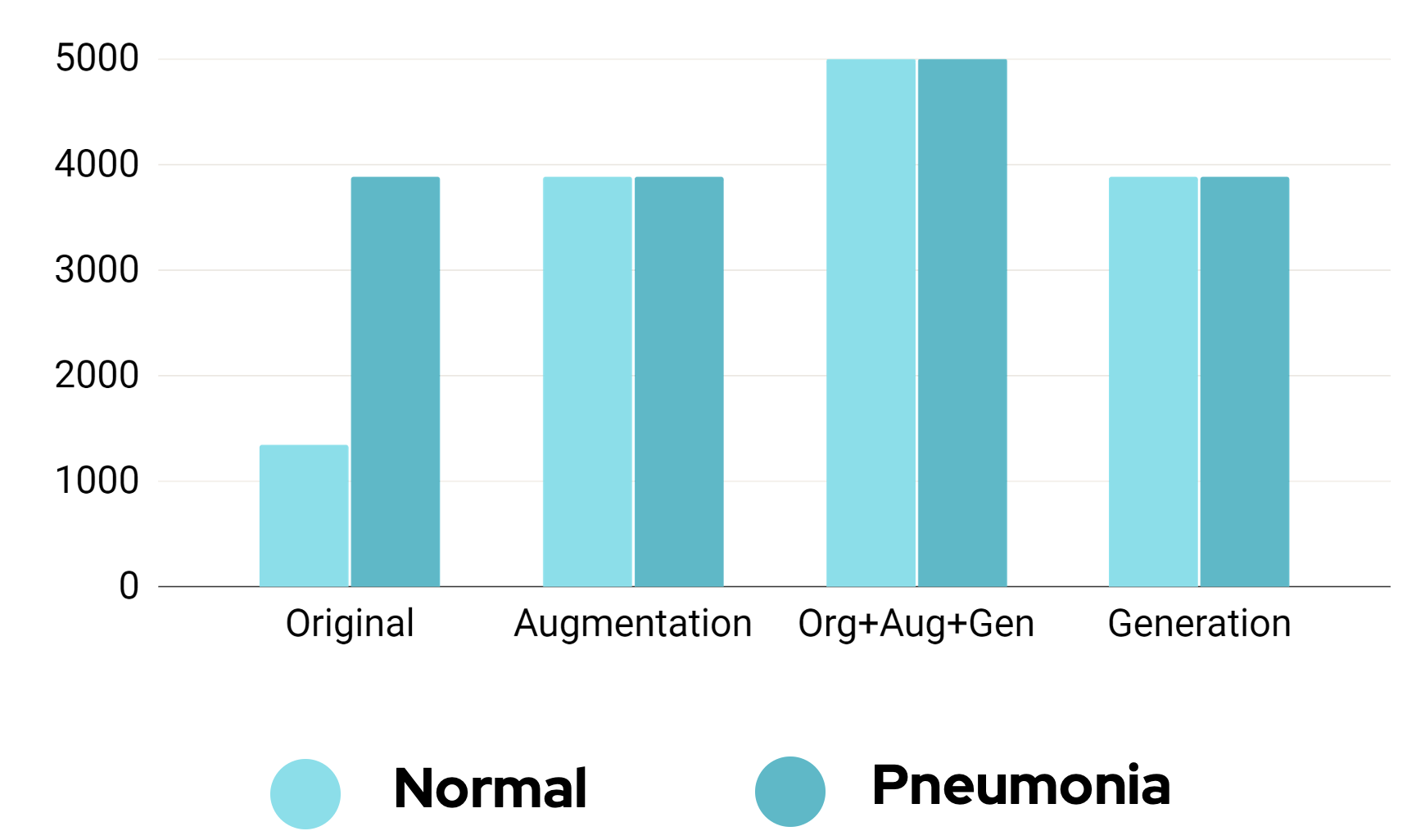}
    \caption{Distribution of the Dataset Classes: Original, Augmented, and GAN-Generated Images.}
    \label{fig:dataset_classes}
\end{figure}

The dataset used in this study is publicly available on Kaggle\footnote{\url{https://www.kaggle.com/datasets/paultimothymooney/chest-xray-pneumonia}} and contains 5,863 chest X-ray images of pediatric patients aged 0 to 5 years. These images were sourced from the Guangzhou Women and Children's Medical Center. The dataset consists of two classes: normal and pneumonia. Each image is labeled by two physicians and verified by a third-party evaluator to ensure accuracy \ref{fig:dataset_classes}. The dataset serves as the foundation for training and testing the machine learning models. A significant challenge in using this dataset is its limited size, which is commonly encountered in medical image classification tasks.

\subsection{Data Preprocessing}
Data preprocessing is a critical step to ensure the images are in the appropriate format for model training. The preprocessing steps applied to the chest X-ray images are as follows:

\begin{itemize}
  \item Resizing: All images were resized to a fixed size of 148x148 pixels to ensure uniformity in the input data for the deep learning models.

  \item Grayscale Conversion: Since the dataset comprises grayscale images, no additional color channels were needed, simplifying the preprocessing pipeline.
  
  \item Normalization: The pixel values of the images were scaled to a range of 0 to 1 by dividing the pixel values by 255, allowing the model to process the data efficiently.
  
  \item Data Splitting: The dataset was split into training (80\%), validation (10\%), and test (10\%) sets to ensure that the model was tested on unseen data and could generalize well.
\end{itemize}

\subsection{Data Augmentation}
Due to the limited size of the dataset, data augmentation techniques were applied to artificially expand the training data. This process helps the model generalize better by exposing it to various image transformations. The augmentation techniques used are:

\begin{figure}[h]
    \centering
    \includegraphics[width=0.8\textwidth]{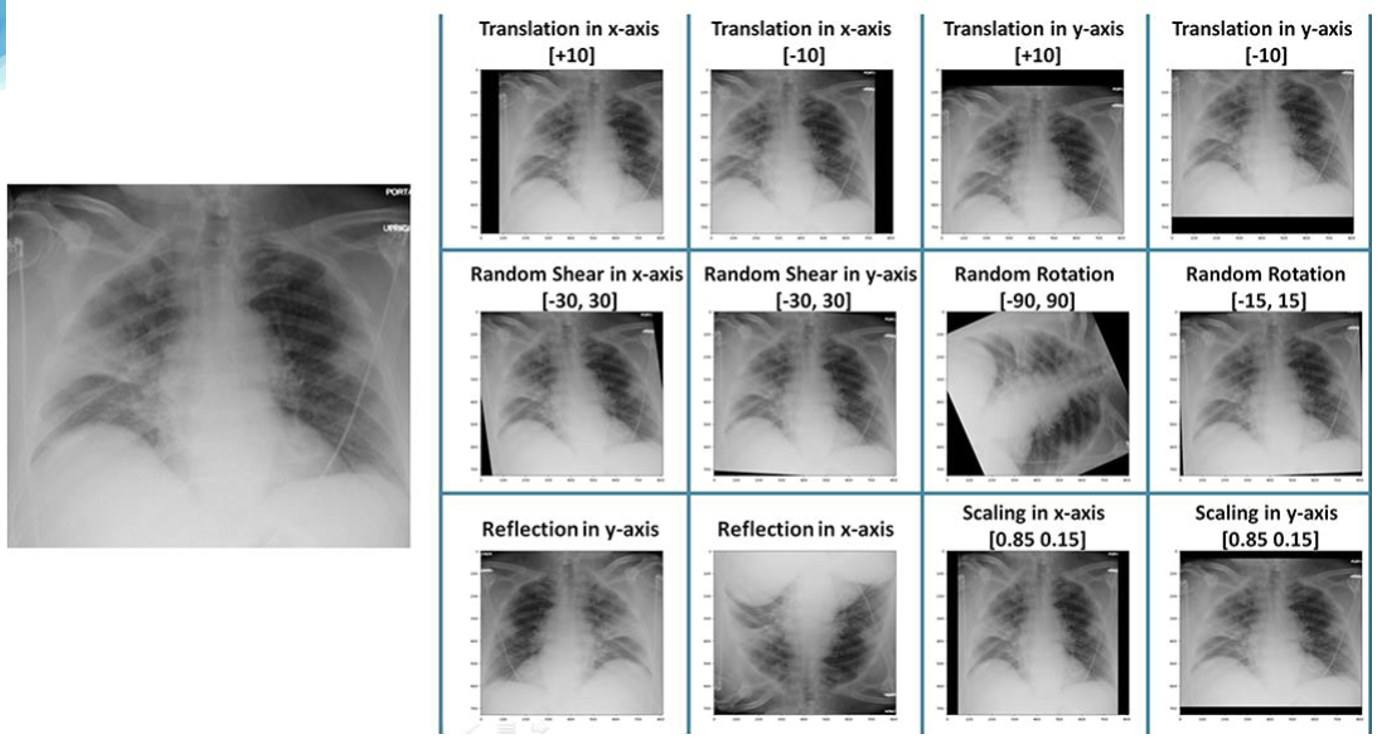}
    \caption{Examples of Augmented Images: Original, Rotated, Zoomed, Sheared, and Flipped.}
    \label{fig:augmented_images}
\end{figure}

\begin{itemize}
  \item Rotation: Random rotations of up to 40 degrees were applied to simulate variations in image angles.

  \item Zoom: Random zooming in and out was used to simulate different distances from the object of interest.

  \item Shear: Random shearing transformations were applied to simulate slight misalignments in images.

  \item Horizontal Flip: The images were randomly flipped horizontally to simulate mirror images.
\end{itemize}

These augmentation techniques increased the diversity of the dataset, helping the model become more robust and reducing the risk of overfitting. An example of the augmented images is shown in Figure \ref{fig:augmented_images}.

\subsection{Generative Adversarial Networks (GANs)}
To further address the problem of limited data, Generative Adversarial Networks (GANs) \cite{DeepLearning2021} were employed to generate synthetic chest X-ray images. GANs consist of two neural networks: the generator and the discriminator. The generator creates new images, while the discriminator evaluates whether the images are real (from the dataset) or fake (generated by the generator). The generator is trained to produce images that are indistinguishable from real images, thereby expanding the dataset and improving model performance. following the architecture shown in Figure \ref{fig:gan_architecture}, the GAN was trained on the original dataset to generate synthetic images that resemble the real chest X-ray images. The GAN architecture consists of:

\begin{figure}[h]
    \centering
    \includegraphics[width=0.6\textwidth]{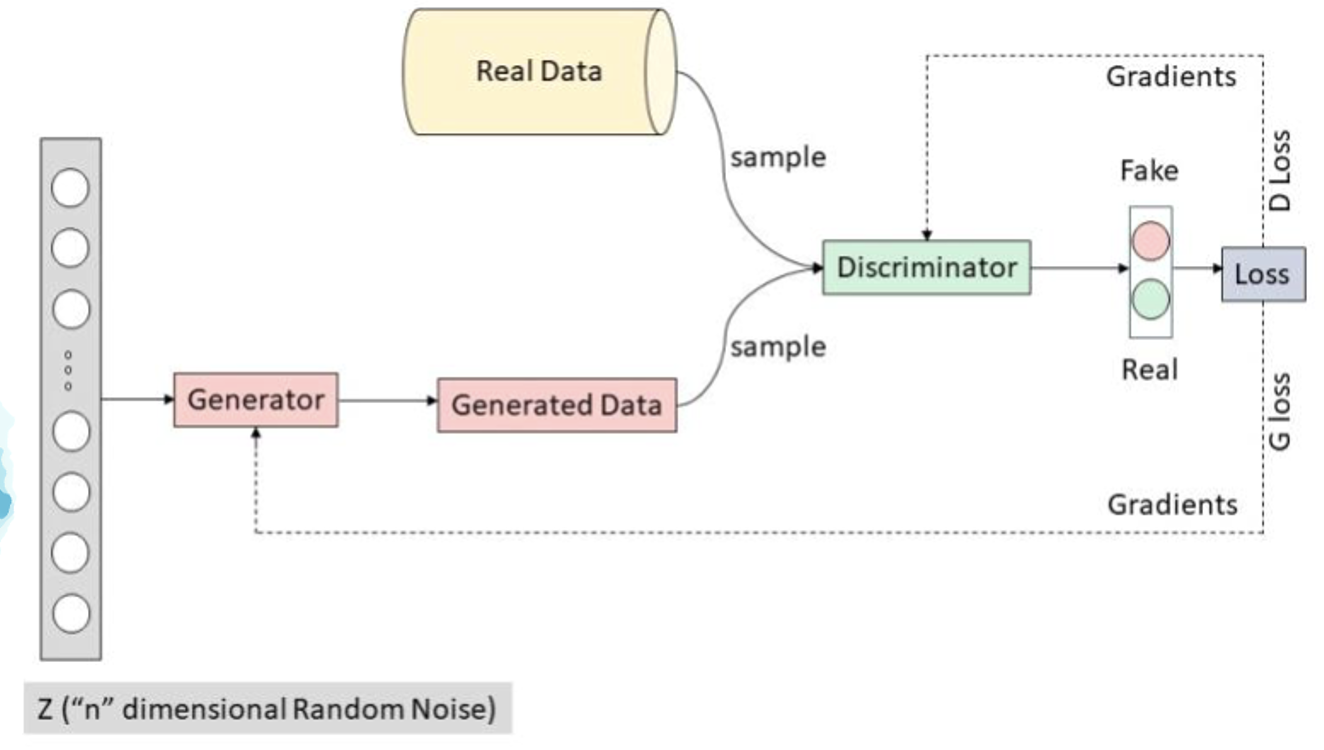}
    \caption{Architecture of the Generative Adversarial Network (GAN) used for generating synthetic chest X-ray images.}
    \label{fig:gan_architecture}
\end{figure}

\begin{figure}[h]
    \centering
    \includegraphics[width=0.8\textwidth]{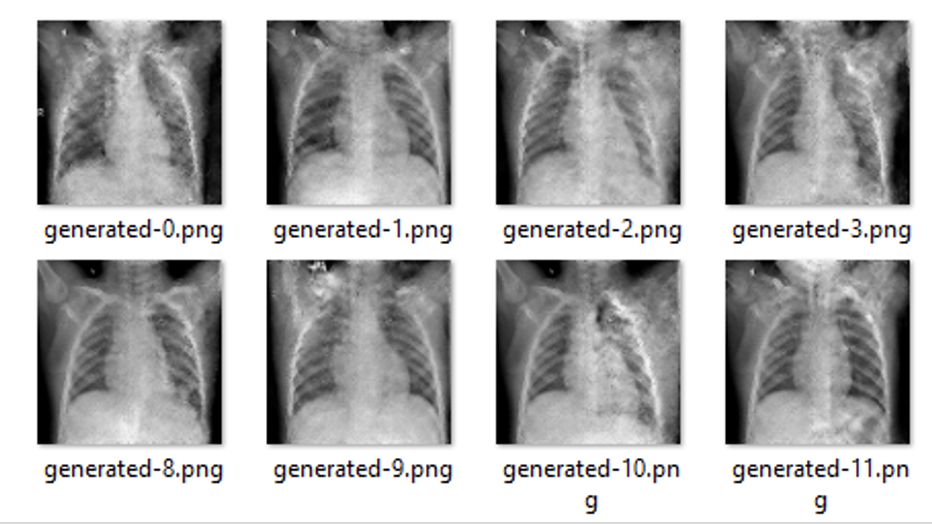}
    \caption{Examples of GAN-Generated Images: Synthetic Chest X-ray Images.}
    \label{fig:gan_samples}
\end{figure}

The generator model was trained to produce 148×148 pixel images, and the GAN was trained over 40,000 epochs to ensure the generated images were of high quality. The synthetic images generated by the GAN were then combined with the real images to create a more balanced and diverse training dataset. The GAN-generated images were evaluated for quality and diversity, ensuring they contributed positively to the training process. An example of the GAN-generated images is shown in Figure \ref{fig:gan_samples}.

\subsection{Convolutional Neural Networks (CNNs)}
The core model used for classification is a Convolutional Neural Network (CNN) \cite{OShea2015}. CNNs are well-suited for image recognition tasks due to their ability to automatically learn spatial hierarchies of features from the input data. The architecture \ref{fig:cnn_architecture} of the CNN used in this study includes:

\begin{figure}[h]
    \centering
    \includegraphics[width=0.5\textwidth]{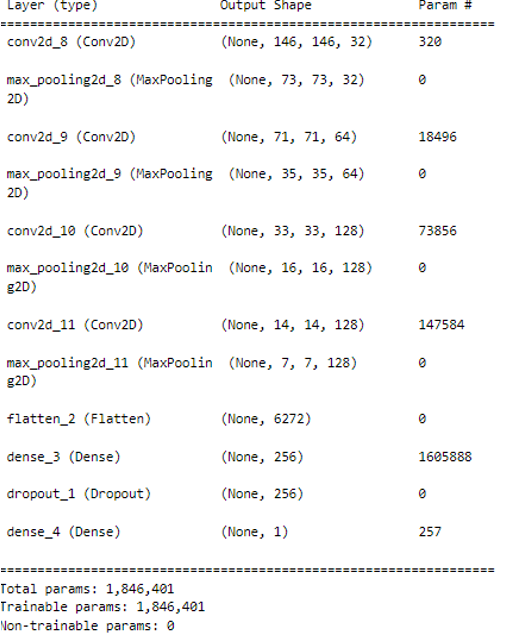}
    \caption{Architecture of the Convolutional Neural Network (CNN) used for Pediatric Chest Pneumonia Classification.}
    \label{fig:cnn_architecture}
\end{figure}

\begin{itemize}
  \item Convolutional Layers: Multiple convolutional layers were used to extract features from the images, with increasing filter sizes (32 → 64) at each layer to capture finer details.

  \item Max-Pooling Layers: Max-pooling layers were added to reduce the dimensionality of the feature maps and retain the most important features.

  \item Batch Normalization: This technique was applied to normalize the outputs of each layer, improving the stability and speed of training.

  \item Fully Connected Layer: The final layer was a fully connected layer with a single output neuron using a sigmoid activation function to classify images as either normal or pneumonia.
  
\end{itemize}

The model was trained using the Adam optimizer with a binary cross-entropy loss function, making it suitable for binary classification. The training process also included early stopping to prevent overfitting and ensure optimal performance.

\subsection{Evaluation Metrics}
The performance of the classification model was evaluated using several metrics:

\begin{itemize}
  \item \textbf{Accuracy:} The proportion of correctly classified images out of the total number of images.
  \[
  \text{Accuracy} = \frac{TP + TN}{TP + TN + FP + FN}
  \]
  where \( TP \) = True Positives, \( TN \) = True Negatives, \( FP \) = False Positives, and \( FN \) = False Negatives.

  \item \textbf{Precision:} The ratio of true positive predictions to the total predicted positives, indicating the model's ability to identify pneumonia cases accurately.
  \[
  \text{Precision} = \frac{TP}{TP + FP}
  \]
  
  \item \textbf{Recall:} The ratio of true positive predictions to the actual positives, measuring the model's ability to detect all pneumonia cases.
  \[
  \text{Recall} = \frac{TP}{TP + FN}
  \]
  
  \item \textbf{F1-Score:} The harmonic mean of precision and recall, providing a balanced measure of the model's performance, especially in cases of class imbalance.
  \[
  \text{F1-Score} = 2 \cdot \frac{\text{Precision} \cdot \text{Recall}}{\text{Precision} + \text{Recall}}
  \]

\end{itemize}

\subsubsection{Experimental Setup}
The experiments were conducted using Python with TensorFlow and Keras libraries. The model was trained on a machine with an NVIDIA GPU to accelerate the training process. The training was performed over multiple epochs, with early stopping implemented to prevent overfitting. The model's performance was evaluated on the validation set after each epoch, and the best model was selected based on the validation accuracy. following the experimental setup, four different experiments were conducted to evaluate the impact of data augmentation and GAN-generated images on model performance. The distribution of classes in each experiment is summarized in Table \ref{tab:experimental_setup}.

\begin{table}[h!]
  \centering
 \begin{tabular}{|c|c|c|c|c|c|}
\hline
\textbf{Experiments} & \textbf{Classes} & \textbf{Original} & \textbf{Augmentation} & \textbf{Generation} & \textbf{Total} \\
\hline
\multirow{2}{*}{1} & Normal & 1349 & 0 & 0 & 1349 \\
\cline{2-6}
 & Pneumonia & 3883 & 0 & 0 & 3883 \\
\hline
\multirow{2}{*}{2} & Normal & 1349 & 2534 & 0 & 3883 \\
\cline{2-6}
 & Pneumonia & 3883 & 0 & 0 & 3883 \\
\hline
\multirow{2}{*}{3} & Normal & 1349 & 0 & 2534 & 3883 \\
\cline{2-6}
 & Pneumonia & 3883 & 0 & 0 & 3883 \\
\hline
\multirow{2}{*}{4} & Normal & 1349 & 1825 & 1826 & 5000 \\
\cline{2-6}
 & Pneumonia & 3883 & 1117 & 0 & 5000 \\
\hline
\end{tabular}
  \caption{Summary of Experimental Setup: Distribution of Classes in Each Experiment.}
  \label{tab:experimental_setup}
\end{table}

\section{Results}

The results presented in this section reflect the performance of the Pediatric Chest Pneumonia Classification System, focusing on the CNN model's accuracy, the influence of data augmentation and GAN-generated images, and the model's deployment in a web application.

The performance of the CNN model was evaluated using several key metrics, including Accuracy, Precision, Recall, and F1-Score. These metrics were measured for four different experiments: the original dataset, augmented data, GAN-generated images, and a combination of original, augmented, and generated images. The results are summarized in Table~\ref{tab:performance_metrics}.

\begin{table}[h]
  \centering
\begin{tabular}{|c|c|c|c|c|}
\hline
 & \textbf{Accuracy} & \textbf{Precision} & \textbf{Recall} & \textbf{F1-Score} \\
\hline
\textbf{Original} & 0.76 & 0.72 & 1.00 & 0.84 \\
\hline
\textbf{Augmented} & 0.80 & 0.76 & 0.98 & 0.86 \\
\hline
\textbf{Generated} & 0.84 & 0.85 & 0.89 & 0.87 \\
\hline
\textbf{Org+Aug+Gen} & 0.86 & 0.84 & 0.95 & 0.89 \\
\hline
\end{tabular}
  \caption{Performance Metrics of the CNN Model for Each Experiment.}
  \label{tab:performance_metrics}
\end{table}

\begin{figure}[h]
    \centering
    \begin{subfigure}[b]{0.48\textwidth}
        \centering
        \includegraphics[width=\textwidth]{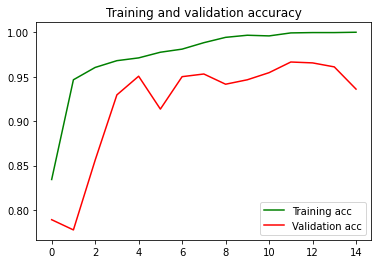}
        \caption{Training and Validation Accuracy of the CNN Model Over Epochs.}
        \label{fig:training_accuracy}
    \end{subfigure}
    \hfill
    \begin{subfigure}[b]{0.48\textwidth}
        \centering
        \includegraphics[width=\textwidth]{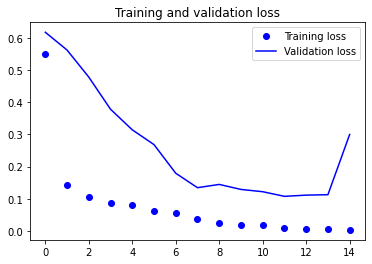}
        \caption{Training and Validation Loss of the CNN Model Over Epochs.}
        \label{fig:training_loss}
    \end{subfigure}
    \caption{Training and Validation Accuracy and Loss of the CNN Model Over Epochs.}
    \label{fig:training_accuracy_loss}
\end{figure}

Figures~\ref{fig:training_accuracy_loss} present the training accuracy and training loss of the CNN model over the epochs. The subfigures include both the accuracy and loss to provide a comprehensive view of the model's performance during training.

\begin{figure}[h]
    \centering
    \includegraphics[width=0.5\textwidth]{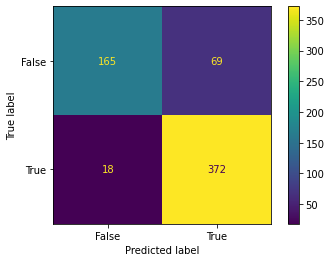}
    \caption{Confusion Matrix of the CNN Model: True Positives, True Negatives, False Positives, and False Negatives.}
    \label{fig:confusion_matrix}
\end{figure}

Figure~\ref{fig:confusion_matrix} displays the confusion matrix of the CNN model, which highlights the distribution of predictions in terms of true positives, true negatives, false positives, and false negatives. The confusion matrix provides insights into the model's performance across the two classes (Normal and Pneumonia). The matrix indicates that the combined model performed exceptionally well in classifying both classes, with a significant number of true positives and true negatives.

As shown in the results, the use of data augmentation and GAN-generated images significantly improved the performance of the CNN model. The GAN-generated images helped balance the dataset, addressing class imbalance and providing the model with more varied data. Combined with the augmentation techniques, this allowed the model to generalize better and achieve higher performance across all metrics.

The combined dataset (original, augmented, and generated) resulted in the best model performance, with improvements in accuracy, precision, recall, and F1-score. The integration of GANs and data augmentation proved to be effective strategies in enhancing model performance, making the system more robust and reliable for real-world applications. The model's deployment in a web application with a user-friendly interface further facilitates its practical use in healthcare settings, enabling timely and accurate diagnoses of pediatric pneumonia from chest X-rays.

\section{Deployment}
The developed Pediatric Chest Pneumonia Classification System was deployed as a web application using Flask, a lightweight web framework for Python. The web application provides an intuitive interface for healthcare professionals to upload chest X-ray images and receive real-time classification results. The deployment process involved the following steps:
\begin{figure}[h]
    \centering
    \includegraphics[width=0.8\textwidth]{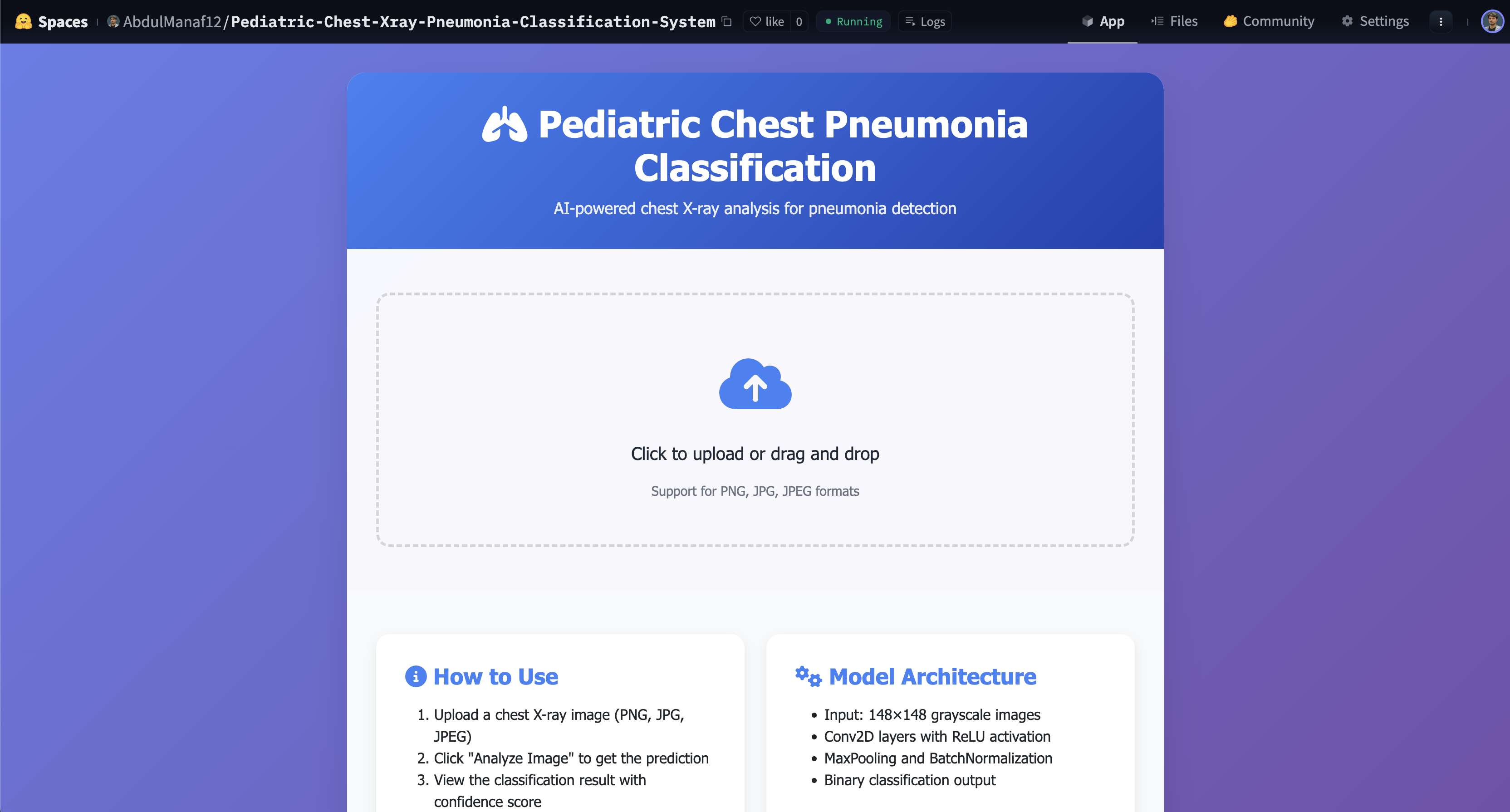}
    \caption{Web Application Interface for Pediatric Chest Pneumonia Classification System.}
    \label{fig:web_application_demo}
\end{figure}

\begin{figure}[h]
    \centering
    \includegraphics[width=0.8\textwidth]{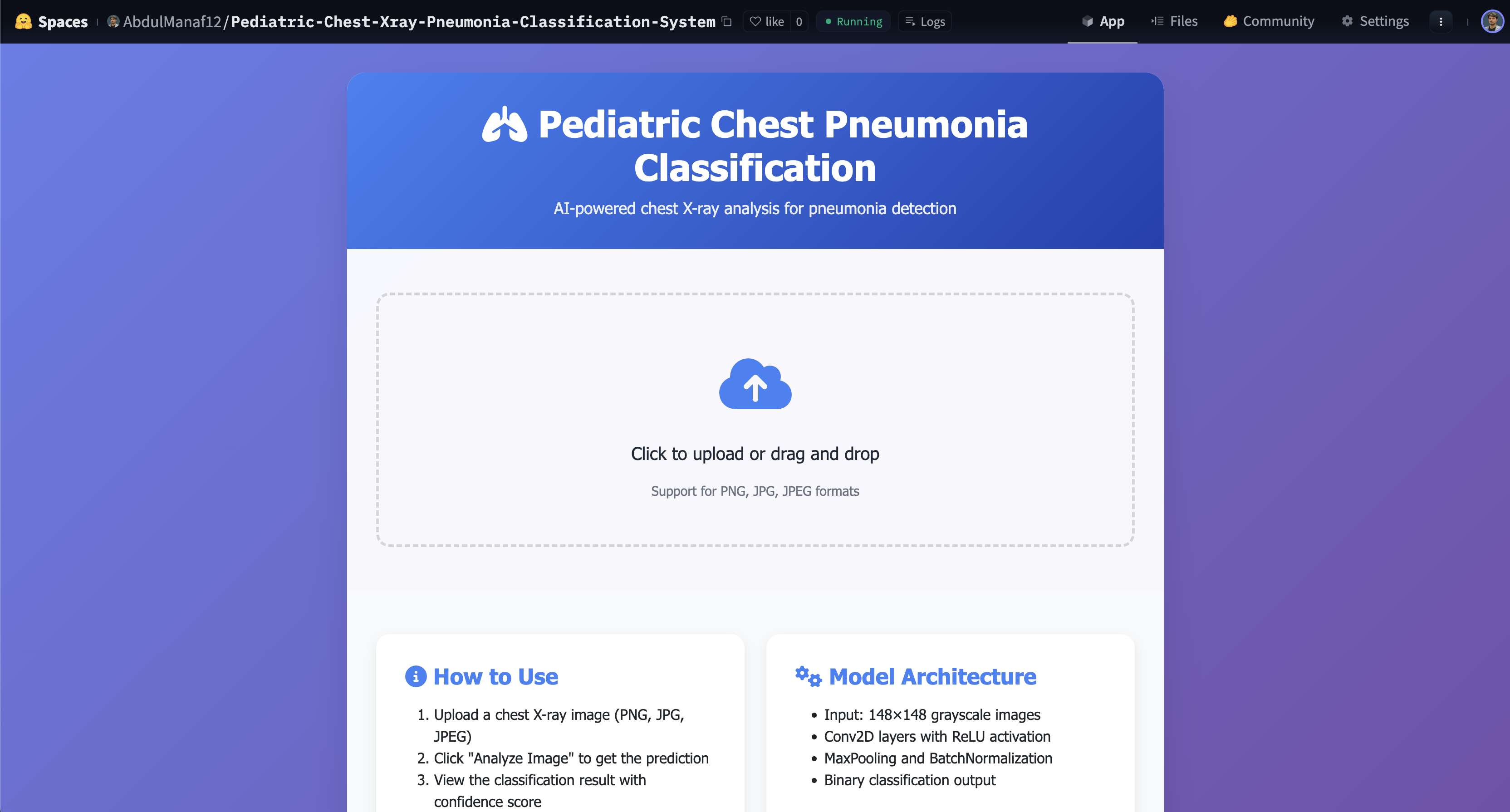}
    \caption{Classification Result from the Web Application: Predicted Class and Probability Score.}
    \label{fig:web_application_result}
\end{figure}

The web application allows users to upload chest X-ray images in JPEG format. Upon submission, the image is processed by the trained CNN model, which classifies it as either "Normal" or "Pneumonia." The application also provides a probability score indicating the model's confidence in its prediction. The user interface is designed to be simple and user-friendly, enabling healthcare professionals to quickly and easily obtain diagnostic results. The web application can be accessed at \url{https://huggingface.co/spaces/AbdulManaf12/Pediatric-Chest-Xray-Pneumonia-Classification-System}. Figure \ref{fig:web_application_demo} shows the web application interface, where users can upload chest X-ray images for classification. After the image is uploaded, the model processes it and returns the classification result along with a probability score. Figure \ref{fig:web_application_result} illustrates the classification result displayed in the web application, showing the predicted class (Normal or Pneumonia) and the corresponding probability score. This feature allows healthcare professionals to make informed decisions based on the model's predictions and provides transparency regarding the model's confidence in its classification. The web application serves as a practical tool for healthcare professionals, enabling them to leverage AI technology for real-time pneumonia classification from chest X-ray images. The deployment of the system in a web application format enhances accessibility and usability, allowing healthcare professionals to utilize the AI-based classification system in clinical settings without requiring extensive technical expertise.

\section{Conclusion}
This study successfully developed a machine learning-based Pediatric Chest Pneumonia Classification System using Convolutional Neural Networks (CNNs) and Generative Adversarial Networks (GANs). The system was trained on a dataset of 5,863 chest X-ray images of children aged 0-5 years, addressing the challenges of limited data through advanced data augmentation techniques and GAN-generated synthetic images. The model achieved high performance metrics, including accuracy, precision, recall, and F1-score, demonstrating its potential to assist healthcare professionals in diagnosing pneumonia from chest X-ray images.
Additionally, Future work could explore the integration of more advanced deep learning architectures, such as transfer learning with pre-trained models, to further enhance classification accuracy.

\bibliographystyle{unsrt}
\bibliography{references}

\end{document}